\documentclass[aps,pre,reprint,superscriptaddress,doi=false,isbn=false,url=false,title=true,longbibliography,noeprint,footinbib]{revtex4-1}
\usepackage{graphicx}
\usepackage{amsmath}
\usepackage{array}
\usepackage{amsfonts}
\usepackage{float}
\usepackage{xcolor}
\def\ba{\begin{eqnarray}}
\def\ea{\end{eqnarray}}
\def\be{\begin{equation}}
\def\ee{\end{equation}}
\def\bm{\begin{math}}
\def\me{\end{math}}

\def\del{\partial}

\newcommand{\dummy}

\makeatletter
\newcommand{\fmarki}{*}
\newcommand{\fmarkii}{\ensuremath{\dagger}}
\newcommand{\fmarkiii}{\ensuremath{\ddagger}}
\newcommand{\fmarkiv}{\ensuremath{\mathsection}}
\newcommand{\fmarkv}{\ensuremath{\mathparagraph}}
\newcommand{\fmarkvi}{\ensuremath{\|}}
\newcommand{\fmarkvii}{**}
\newcommand{\fmarkviii}{\ensuremath{\dagger\dagger}}
\newcommand{\fmarkix}{\ensuremath{\ddagger\ddagger}}
                
\def\@fnsymbol#1{{\ifcase#1\or \fmarki\or \fmarkii\or \fmarkiii\or \fmarkiv\or \fmarkv\or \fmarkvi\or \fmarkvii\or \fmarkviii\or \fmarkix \else\@ctrerr\fi}}
\makeatother

\begin{document}

\title{Activity Induced Enhanced Diffusion of a Polymer in Poor Solvent}
\author{Suman Majumder}\email[]{ suman.jdv@gmail.com}
\affiliation{Amity Institute of Applied Sciences, Amity University Uttar Pradesh, Noida 201313,
India
}
\author{Subhajit Paul}\email[]{ subhajit.paul@icts.res.in}
\affiliation{International Center for Theoretical Sciences, Tata Institute of Fundamental Research, Bangalore-560089, India}

\author{Wolfhard Janke}\email[]{ wolfhard.janke@itp.uni-leipzig.de}
\affiliation{Institut f\"{u}r Theoretische Physik, Universit\"{a}t Leipzig, IPF 231101, 04081 Leipzig, Germany 
}




\date{\today}

\begin{abstract}
By means of Brownian dynamics simulations we study the steady-state dynamic properties of a flexible active polymer in  a poor solvent condition. Our  results show that the effective diffusion constant of the polymer $D_{\rm eff}$ gets significantly enhanced as activity increases, much like in active particles. The simulation data are in agreement with a theoretically constructed Rouse model of active polymer, demonstrating that irrespective of the strength of activity, the long-time dynamics of the polymer chain is characterized by a universal Rouse-like scaling $D_{\rm eff} \sim N^{-1}$, where $N$ is the chain length. 
\end{abstract}


\maketitle

Biomolecules are subjected to athermal fluctuations originating from chemical reactions or other energy conversions, rendering them fall out of equilibrium. Often that is the underlying cause for a range of biological activities, e.g., bacterial motion, shape fluctuations of red-blood cell membranes, enzyme catalysis \cite{Wu2000,park2010,ben2011,jee2018,jee2018enzyme}. Hence, given the enormous progress in understanding of active particles \cite{Ramaswamy2010,shaebani2020,elgeti2015}, over the years a number of studies have emerged investigating active polymers as well \cite{winkler2020,kaiser2015,isele2015,isele2016,bianco2018,locatelli2021,das2021,paul2021,paul2022,paul2022role,paul2022activity}. The motivation for studying active polymers stems from the need of introducing variety in shape, flexibility and coupling topology in active entities \cite{paxton2004,sanchez2011,ebbens2012,buttinoni2013,kummel2013,zottl2014}. Besides, it is intriguing to check how the knowledge of polymer physics  \cite{de1979scaling,doi1996,rubinstein2003} can be deployed to understand active matters.
\par
To date active polymer models can be classified into two categories. The straightforward way is to consider the monomers as active particles and then connecting them linearly with a bond constraint \cite{kaiser2015,bianco2018,das2021,paul2021,paul2022,paul2022role,paul2022activity}. In the other approach one takes a  passive polymer, i.e., without any activity, immersed in a bath of active particles \cite{harder2014,kaiser2014,chaki2019,liu2019,anderson2022,goswami2022}. Apart from being motivated by biological entities, current advanced techniques allows one to   
synthesize polymers made of active colloids connected artificially by DNA or freely jointed droplets  \cite{biswas2017,mcmullen2018}. Theoretically, the constituent monomers can be made active by (\textit{i}) introducing a local
force tangential to the polymer backbone \cite{bianco2018}, (\textit{ii})   considering the monomers having Brownian activity \cite{das2021,paul2022activity}, or (\textit{iii}) Vicsek-like activity  \cite{paul2021,paul2022,paul2022role}.
\par
The conformational and dynamic properties of a passive polymer is characterized by various well established scaling laws \cite{de1979scaling,doi1996,rubinstein2003}. A polymer undergoes a coil to globule transition upon changing the solvent condition from good (where monomer-solvent interaction dominates) to poor (where monomer-monomer interaction dominates). The spatial extension of the conformations measured in terms of the radius of gyration $R_g$ typically follows the scaling $R_g \sim N^{\nu}$ with respect to the degree of polymerization or number of monomers $N$. The value of the exponent $\nu \approx 3/5$ and $1/3$ characterizes the conformations, respectively, in good and bad solvents. The dynamics of a polymer under a good solvent condition in the free-draining limit, i.e., ignoring hydrodynamics, is characterized by Rouse  scaling $D \sim N^{-1}$, where $D$ is the diffusion coefficient of the center of mass of the polymer \cite{rouse1953}. In presence of hydrodynamics, one expects  in the Zimm model \cite{zimm1956} (with excluded volume) a scaling $D\sim N^{-\nu}$. In a poor solvent condition, however, there is no consensus among the available studies \cite{naghizadeh1987,milchev1993}. Anomalous behavior with resemblance to slow glassy dynamics has also been reported \cite{milchev1994}. 
\par
In case of active polymers too, the focus has been on understanding the nonequilibrium steady-state conformations and dynamics. In particular, attempts have been made to adapt scaling theories of passive polymers in good solvent to study active polymers under the same condition. Recently, Bianco \textit{et al.} \cite{bianco2018} observed an activity induced collapse of a polymer in good solvent, reminiscent of motility induced phase separation of active particles \cite{fily2012}. In contrast, polymers made of active Brownian monomers do not exhibit such collapse in good solvent. On the dynamics front, in steady state an enhancement of the diffusion coefficient has been observed for all cases. In all these studies the self-avoidance in good solvent condition has been mimicked by considering a purely repulsive interaction among the constituent monomers. Only recently, an interaction potential with both attractive and repulsive components has been considered \cite{das2021,paul2021,paul2022,paul2022role,paul2022activity}. A passive polymer having such an interaction exhibits a temperature dependent coil-globule transition.  We have shown that a polymer with active Brownian monomers in bad solvent condition, exhibits a transition from a globular state at small activity to coil-like conformations at large activity \cite{paul2022activity}. Hence, it is expected that the dynamics of such active polymers in poor solvent would also reveal interesting features.
\par
In this Letter, by means of computer simulation supported by analytical reasoning we investigate the steady-state dynamics of a flexible coarse-grained model polymer consisting of active Brownian monomers in a poor solvent condition. To probe the dynamics we have monitored the motions of the center of mass of the polymer and two different tagged monomers, viz., the central and end monomers. All these motions exhibit long-time diffusive behaviors allowing us to calculate the diffusion coefficient $D$ of the polymer. We show that a universal Rouse-like scaling of $D$ with respect to $N$ is maintained at all considered strengths of activity, albeit, the polymer could be in a globular or coiled state.

We consider a flexible polymer consisting of monomers of diameter $\sigma$ at positions $\vec{r}_i$ connected linearly via spring-like bonds. Its dynamics is governed by the over-damped Langevin equations 

\begin{equation}\label{trans}
\begin{split}
\del_t{\vec{r}}_i &=\frac{ D_{\rm{tr}}}{k_BT}[f_p \hat{n}_i-\vec{\nabla} U_i]+\sqrt{2D_{\rm{tr}}}\,\vec{\Lambda}_i^{\rm{tr}}, \\
\del_t{\hat{n}}_i &= \sqrt{2D_{\rm{rot}}}(\hat{n}_i\times \vec{\Lambda}_i^{\rm{rot}}),
\end{split}
\end{equation} 
where $f_p$ is the strength of the self-propulsion force acting along the unit vector $\hat{n}_i$ that changes stochastically with time, and $U_i=V_{\rm{FENE}}+V_{\rm{LJ}};$ where $V_{\rm{FENE}}(r) = -0.5 KR^2 {\rm{ln}}[ 1- ((r-r_0)/R)^2]$ takes care of the bonds between successive monomers and $
V_{\rm{LJ}}(r) = 4\epsilon [(\sigma/r)^{12}- (\sigma/r)^6]$ is the non-bonded interaction accounting for the self-avoidance, with a strength $\epsilon$ . In Eq.\ \eqref{trans}  $D_{\rm{tr}}$ and $D_{\rm{rot}}$ are, respectively, the translational and rotational diffusion constants. The components of
 $\vec{\Lambda}_i^{\rm{tr}}$ and $\vec{\Lambda}_i^{\rm{rot}}$ are independent white Gaussian noises with zero-mean and unit-variance. In our simulations we have fixed their relative importance
	$D_{\rm{tr}}/D_{\rm{rot}}\sigma^2\equiv \Delta =1/3$.
 $D_{\rm{tr}}$ is related to the temperature $T$ and the drag or friction coefficient $\gamma$ as $D_{\rm{tr}}=k_BT/\gamma$. For convenience we have chosen $\gamma=1$. Time is measured in units of $\tau_0=\sigma^2\gamma/\epsilon$ ($\propto 1/D_{\rm{rot}}=\Delta \sigma^2 \gamma /k_BT$ at fixed $k_BT/\epsilon$) and we have used an  integration time step of $ 10^{-5}\tau_0$. Here onward, the activity strength $f_p$ will be expressed in terms of the dimensionless P\'eclet number $Pe={f_p\sigma}/{k_B T}$.
The case of $Pe=0$ corresponds to a passive polymer. We perform our simulations for different $Pe$ at $T=0.1\epsilon/k_B$,  well below the 
coil-globule transition temperature of the passive polymer \cite{majumder2017}, thus mimicking a poor solvent condition. For further technical details on the model and simulation method we refer to  the Supplemental Material \footnote{See Supplemental Material at [\dots] which includes Refs.\ \cite{rouse1953,howse2007}. The Supplemental Material contains a detailed description of the model and method of simulation used, and a few representative steady-state trajectories of the polymer. It also includes details of the analogous Rouse model for active polymer in poor solvent, and a derivation of the corresponding diffusion constant of the center of mass. Furthermore, we present details of the performed fitting exercise showing the agreement between theory and simulation data.
}\label{Supple}.

\par

\begin{figure}[t!]
\centering
\includegraphics*[width=0.45\textwidth]{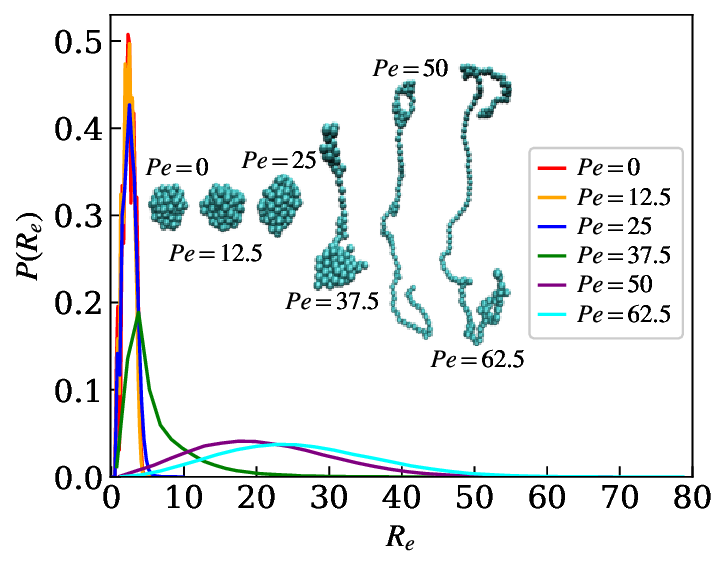}
\caption{\label{snapshots}  Typical steady-state conformations of a polymer of length $N=128$ at different activity strength $Pe$, obtained from simulations at a fixed temperature $T=0.1 \epsilon/k_B$. The plots are for corresponding normalized distributions of the end-to-end distance $R_e$.}
\end{figure}

We started our simulations using self-avoiding coils as initial condition. Then, we allow  the system to reach its steady state by running it for sufficiently long times. Typical steady-state conformations of a polymer of length $N=128$ obtained 
for different $Pe$ are presented in Fig.\ \ref{snapshots}. The conformation of the passive polymer, i.e., $Pe=0$, is a perfectly collapsed globule. It remains in such a globular state for relatively smaller activities $Pe \le 25$ as well. For intermediate values of $Pe$ one may observe a globule or  head-tail-like conformations (as the one presented for $Pe=37.5$). For even larger $Pe$ the polymer becomes an extended coil. The corresponding quantitative picture is presented in Fig.\ \ref{snapshots} in the form of the distribution of end-to-end distance $R_e= |\vec{r}_1-\vec{r}_N|$ where $\vec{r}_i$ is the position of the $i$-th monomer. It shows that for $Pe < 37.5$ the peaks are at $R_e \approx 3$, indicative of   collapsed globules. The decrease in peak height as $Pe$ increases is reflective of the fact that the probability of getting a collapsed globule decreases and encountering  a head-tail-like conformations increases. 
For $Pe > 37.5$ the distribution broadens and the peak position shifts towards $R_e > 15$ suggesting a dominance of coil-like conformations. The overall picture is reminiscent of the temperature driven coil-globule transition of a passive polymer. Here, it is driven by the activity strength.  In fact, we have confirmed in Ref.\ \cite{paul2022activity} that  the conformations obey the scaling law $R_g \sim N^\nu$ with $\nu=1/3$ and $\approx 3/5$, respectively, at small and large $Pe$.
\begin{figure*}[t!]
\centering
\includegraphics*[width=0.85\textwidth]{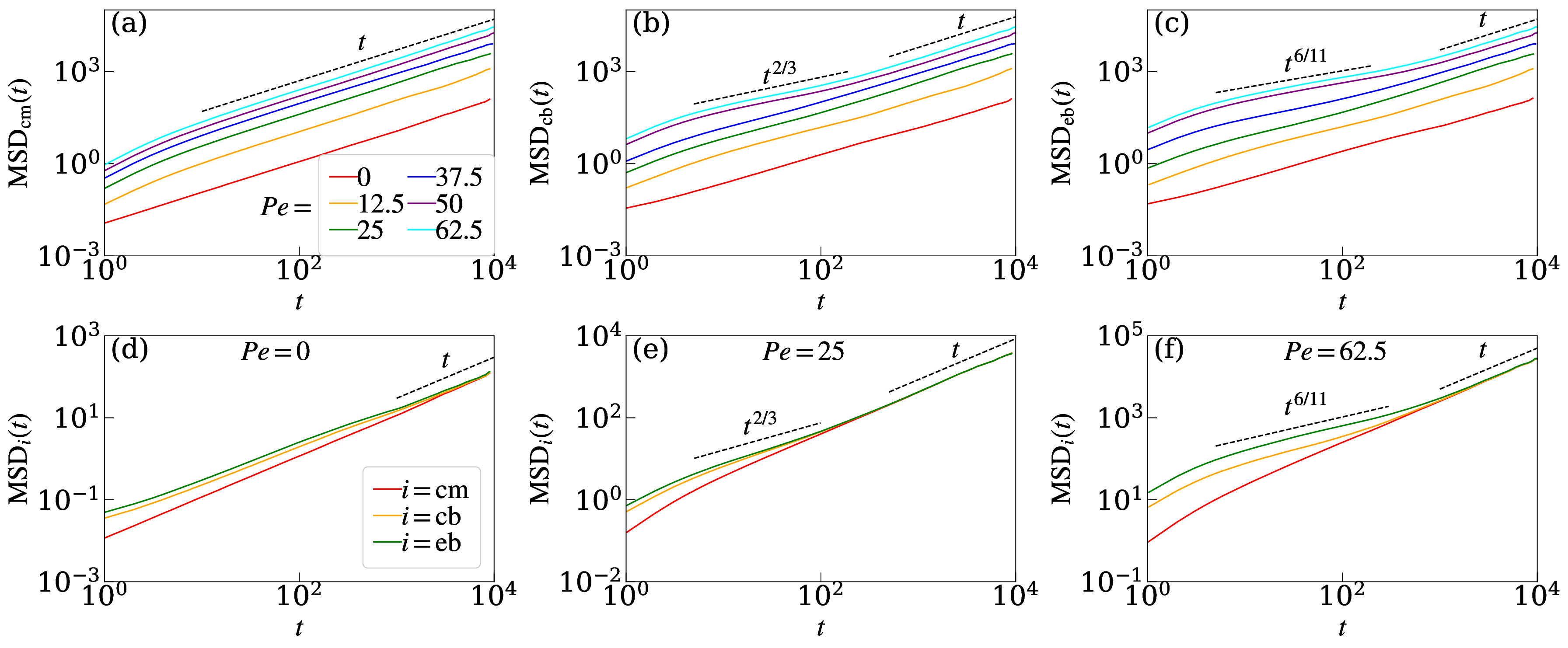}\\
\caption{\label{msd_diff_fa} Steady-state mean square displacement ${\rm MSD}_i(t)$ of the (a) center of mass, (b) central bead, and (c) end beads of a polymer of length $N=128$ for different activity strengths $Pe$. Plots in (d), (e), and (f) present a comparison among the different MSDs for different values of $Pe$. The dashed lines represent different power laws. All data are obtained from simulations at temperature $T=0.1\epsilon/ k_B$.}
\end{figure*}
\par
To probe the dynamics we monitor the motion of the center of mass (cm), central monomer or bead (cb) and end beads (eb) of the polymer. A bare look at the typical trajectories over a fixed time period reveals that although  motions are random in general, the distance covered varies significantly for different $Pe$, clearly suggesting a difference in dynamics (see Fig.\ S1 in the Supplemental Material \cite{Note1}). To quantify the differences, from the obtained trajectories, we calculate the corresponding mean square displacements 
\begin{equation}\label{MSD}
 {\rm MSD}_i(t) = \langle \left[\vec{r}_i(t)-\vec{r}_i(0) \right]^2\rangle;~ i\equiv {\rm cm, cb, and~eb},
\end{equation}
as a function of time $t$. Figure\ \ref{msd_diff_fa}(a) shows that the cm exhibits a typical long-time diffusive motion $\sim t$, with pronounced short-time ballistic behavior as the activity increases. A similar long-time behavior is also observed for the motions of cb and eb, shown respectively, in Figs.\ \ref{msd_diff_fa}(b) and (c). Significantly, different is the appearance of an intermediate regime which becomes longer as the activity strength $Pe$ increases. In this regime the behavior of the  central bead appears to be $\sim t^{2/3}$ for large $Pe$, which may lure one to consider it as hydrodynamic Zimm's scaling of a passive polymer in good solvent \cite{paul1991,dunweg1991,dunweg1993,hinczewski2009}. However, this is very unlikely and probably a mere coincidence since our simulations do not preserve hydrodynamics.  The end beads show an extended intermediate regime, although,  the corresponding power-law exponent seems to be smaller than $2/3$. This rather suggests a Rouse-like behavior, expected for a passive polymer in good solvent. 

\begin{figure}[b!]
\centering
\includegraphics*[width=0.45\textwidth]{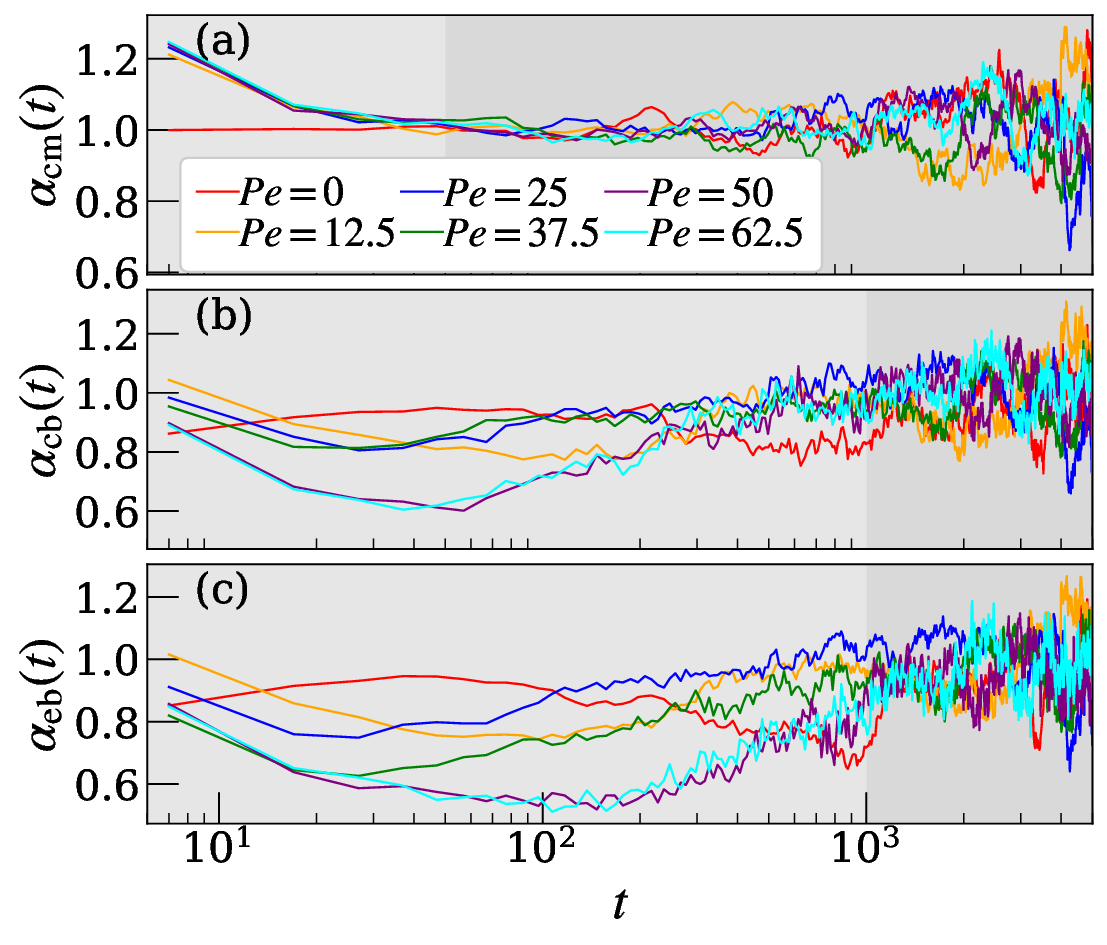}\\
\caption{\label{alpha} Plots of the time-dependent exponent $\alpha_i(t)$ for the data presented in Fig.\ \ref{msd_diff_fa}. The grey shades are introduced to distinguish the early-time regime from the long-time diffusive regime.}
\end{figure}
\par
For a better understanding of the time-dependent power-law behavior ${\rm MSD}_i \sim t^{\alpha}$, we calculate the instantaneous exponent 
\begin{equation}
 \alpha_i(t)= \frac{d\ln {\rm MSD}_i(t)}{d\ln t}; ~ i\equiv {\rm cm, cb, and~eb}.
\end{equation}
Corresponding plots of $\alpha_i(t)$ \textit{vs.} $t$ are presented in Fig.\ \ref{alpha} for all considered values of $Pe$. The exponent $\alpha_{\rm cm}$ for $Pe >0$ starts at a value $ > 1$ and quickly [beginning of the darker shade in Fig.\ \ref{alpha}(a)] approaches $1$, consistent with the long-time diffusive behavior. 
\par
For $Pe \ge 50$, where the polymer is in a coiled state, starting from a value around $0.9$, the exponent $\alpha_{\rm cb}$ drops 
significantly before it climbs up in the diffusive regime [Fig.\ \ref{alpha}(b)]. However, one can hardly see a flat intermediate region to consider this as a true scaling regime. Importantly, the data never really show a steady behavior around the value $2/3$, thus ruling out the apparent Zimm's scaling. This drop in $\alpha_{\rm cb}$ can rather be interpreted as an effect of gradual crossover to the long-time diffusive regime. The crossover gets delayed with increase in $Pe$, as evident  from Figs.\ \ref{msd_diff_fa}(d)-(f) showing that the data for ${\rm MSD}_{\rm cb}(t)$ merge with the one for ${\rm MSD}_{\rm cm}(t)$ at large $t$.
\begin{figure*}[t!]
\centering
\includegraphics*[width=0.85\textwidth]{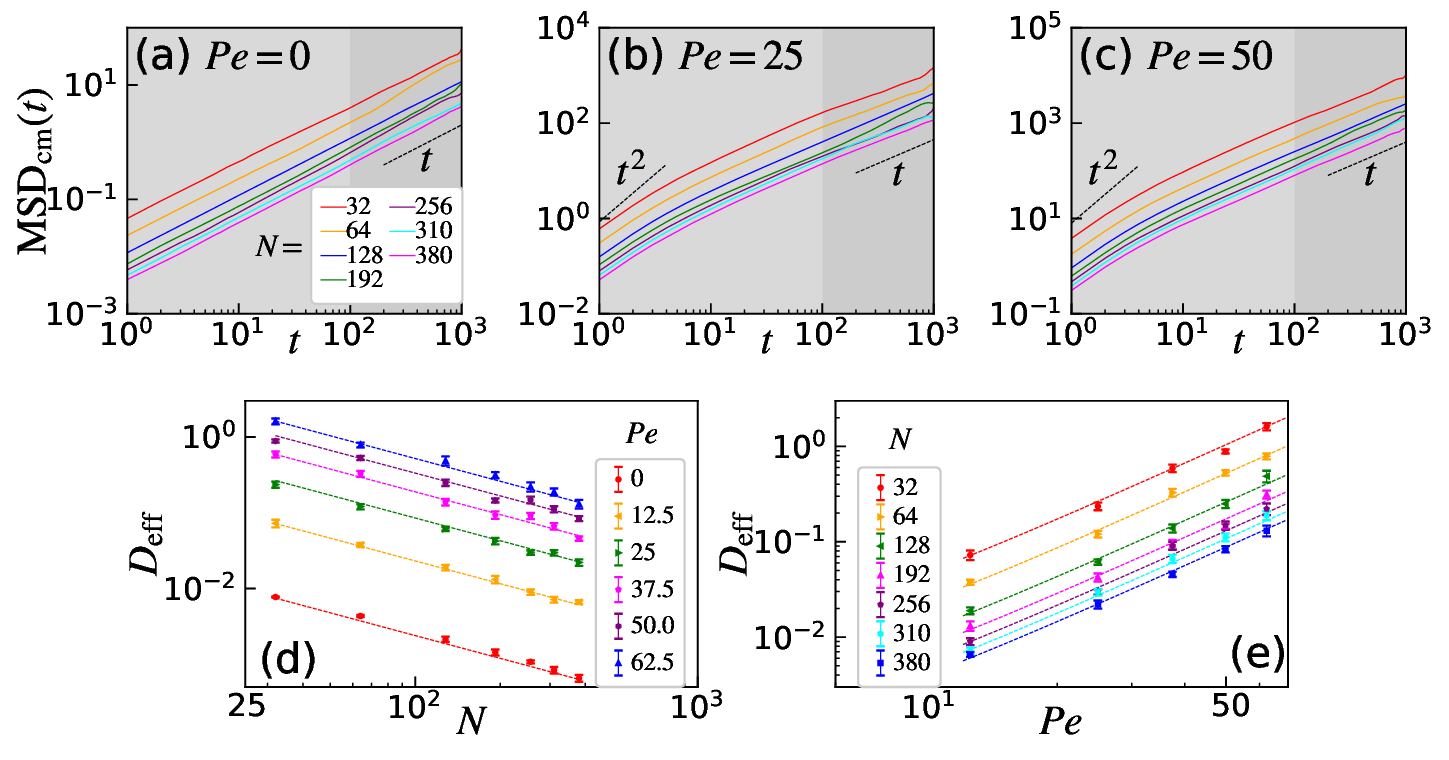}\\
\caption{\label{Diffusion_CM} Chain-length dependence of the mean square displacement of the center of mass at temperature $T=0.1\epsilon/k_B$ for (a) $Pe=0$, (b) $Pe=25$, and (c) $Pe=62.5$. Regions with darker shades mark the time period $t \in [10^2:10^3]$, over which the diffusion coefficient $D_{\rm eff}$ is calculated using Eq.\ \eqref{diff_eq}. (d) Scaling  of $D_{\rm eff}$ with $N$ for different $Pe$. The dashed lines represent the prediction in Eq.\ \eqref{D_eff} with $D_m=0.42$ for fixed $Pe$. (e) $D_{\rm eff}$ as a function of $Pe$ for different $N$. The dashed lines there also represent the prediction in Eq.\ \eqref{D_eff} for fixed $N$.}
\end{figure*}
\par
For the end beads $\alpha_{\rm eb}(t)$ shows a similar behavior of approaching $1$ at late times [Fig.\ \ref{alpha}(c)]. This implies that the data for ${\rm MSD}_{\rm eb}(t)$ must coincide with ${\rm MSD}_{\rm cm}(t)$ at large $t$, which can be verified from the plots in Figs.\ \ref{msd_diff_fa}(d), (e), and (f). Similar to $\alpha_{\rm cb}$, at intermediate times the data for $\alpha_{\rm eb}$ show a drop from $1$ and tend to become flat before it finally approaches $1$ at large $t$. This indicates the presence of a true intermediate power-law regime. For $Pe\ge 50$, the value of $\alpha_{\rm eb}$ in the intermediate flat regime is less than $2/3$. In absence of hydrodynamics effects, i.e., for a Rouse polymer with excluded volume, in the intermediate regime, the scaling for the end monomers is given by  \cite{paul1991,dunweg1991,dunweg1993,shusterman2004} 
\begin{equation}\label{eb_scaling}
{\rm MSD}_{\rm eb}(t) \sim t^{2\nu/(1+2\nu)}.
\end{equation}
For a Gaussian chain having $\nu=1/2$ this provides a $\sim t^{1/2}$ behavior. In the present case at large $Pe$, the polymer behaves like a self-avoiding coil with $\nu \approx 3/5$ producing a scaling $\sim t^{6/11}$. Our data is indeed consistent with such a behavior, shown by the dashed lines in Figs.\ \ref{msd_diff_fa}(c) and (f). Thus it can be inferred that the intermediate Rouse scaling which in general does not hold for a passive polymer in poor solvent \cite{milchev1994}, can be recovered in an active polymer at sufficiently large strength of activity.

\par
For a theoretical understanding, we consider an analog of the Rouse model, the simplest model describing dynamics of a passive polymer in absence of hydrodynamics \cite{rouse1953}.  In the model successive active monomers along the chain are connected via harmonic springs. In addition, each monomer experiences a net random force $\vec{F}_i$ resulting from the thermal noise, active force, and nonbonded interaction in poor solvent condition. Assuming that $\vec{F}_i$ is Delta-correlated over time and space, for the motion of the cm at long time one can write down 
(see details in the Supplemental Material \cite{Note1})
\begin{equation}\label{Rouse_MSD}
\begin{split}
&{\rm MSD}_{\rm cm}(t)=
\langle \left[\vec{r}_{\rm cm}(t)-\vec{r}_{\rm cm}(0) \right]^2\rangle \\
&=
\left\langle \int_0^td{t'} \int_0^td{t'' \left[\frac{1}{N}\sum_{i=1}^{N} \vec{F}_i(t')\right]\cdot 
\left[\frac{1}{N}\sum_{j=1}^{N} \vec{F}_j(t'')\right]}\right\rangle\\
&=6D_{\rm eff}t=\frac{6(D_{\rm a}/D_m)}{N}t.
\end{split}
\end{equation}
Here, $D_{\rm eff}$ is effective diffusion constant of the cm of the polymer which is related to the diffusion constant of the constituent active monomers $D_{\rm a}$  via a modification factor $D_m$, introduced to take into account the poor solvent condition. From the long-time ${\rm MSD}$ of an active Brownian particle \cite{howse2007}, one gets $D_{\rm a}= \left(1+{Pe^2}/{18}\right){k_BT}/{\gamma }$, which on inserting in Eq.\ \eqref{Rouse_MSD} yields (see Supplemental Material \cite{Note1})  
\begin{equation}\label{D_eff}
 D_{\rm eff}=\left(1+\frac{Pe^2}{18} \right)\frac{k_BT}{\gamma D_m N},
\end{equation}
implying a Rouse-like scaling $D_{\rm eff} \sim N^{-1}$ at a fixed $Pe$, and $D_{\rm eff} \sim Pe^2$ for fixed $N$.
\par
To verify the prediction in Eq.\ \eqref{D_eff}, we calculate $D_{\rm eff}$ from our simulation data using the following prescription 
\begin{equation}\label{diff_eq}
 D_{\rm eff}=\frac{1}{6}\lim_{t \rightarrow \infty}\frac{d}{dt}{\rm MSD}_{\rm cm}(t).
\end{equation}
The data of ${\rm MSD}_{\rm cm}(t)$ for polymers of different length $N$, showing a linear behavior in the long-time limit, are  presented in Figs.\ \ref{Diffusion_CM}(a)-(c). The extracted 
 $D_{\rm eff}$ as a function of  $N$ for fixed $Pe$ are presented in Fig.\ \ref{Diffusion_CM}(d). The dashed lines represent Eq.\ \eqref{D_eff} with $D_m=0.42$ which was obtained as the most reasonable choice following a rigorous fitting exercise presented in the Supplemental Material \cite{Note1}. The  consistency of our data with the plotted functions  irrespective of the value of $Pe$ confirms the presence of a universal Rouse-like scaling as embedded in the prediction \eqref{D_eff}. Furthermore, as $Pe$ increases a significant enhancement of $D_{\rm eff}$ can be noticed. Figure \ref{Diffusion_CM}(e) shows this enhancement via plots of $D_{\rm eff}$ as a function $Pe$ for fixed $N$. The functional dependence of $D_{\rm eff}$ on $Pe$ for a fixed $N$ is predicted in Eq.\ \eqref{D_eff}. Fitting the ansatz  (presented in the Supplemental Material \cite{Note1}) using $D_{\rm eff}$ vs.\ $Pe$ data yield the same $D_m=0.42$. Plots of Eq.\ \eqref{D_eff} with $D_m=0.42$, shown by the dashed lines in Fig.\ \ref{Diffusion_CM}(e) not only depict an unambiguous agreement of the prediction  with the simulation data but also indicates that the modification factor $D_m$ is rather universal, independent of $N$ and $Pe$. 

\par
In conclusion, we have presented results for the steady-state dynamics of an active Brownian polymer in poor solvent condition. In order to explore the dynamics we have monitored the motions of the center of mass, central monomer, and the end monomers. The mean square displacement of the end monomers shows the presence of an intermediate regime. In  the large-activity limit, this intermediate regime exhibits a $\sim t^{6/11}$ scaling, which generally holds for Rouse dynamics of a passive polymer in good solvent. In the long-time limit, the mean square displacement of the central and end monomers merge with that of the diffusive behavior of the center of mass. This allows us to estimate the long-time effective diffusion coefficient $D_{\rm eff}$ of the polymer. Analytically, we predict the dependence of $D_{\rm eff}$ on chain length $N$ and activity strength $Pe$ using a Rouse model of active polymer. Our numerical results are in agreement with the theoretical prediction showing a significant enhancement of $D_{\rm eff}$ as a function of $Pe$ obeying a scaling $D_{\rm eff} \sim Pe^2$. Similarly as predicted, the data show that the universal Rouse-like $D_{\rm eff} \sim N^{-1}$ still holds strongly irrespective of $Pe$. It would be interesting to explore the robustness of this Rouse-like behavior for semiflexible polymers with activity \footnote{S. Majumder, S. Paul, and W. Janke, in progress.}. 

\par
This work is the first to explore the steady-state dynamics of an active polymer in poor solvent. As a future endeavour, it would be worth to investigate the same in other active polymer models. Our main result showing activity  induced enhanced diffusion of a polymer in a poor solvent condition might indulge in design of synthetic active polymers which potentially can be employed in delivering drugs  for a wide variety of media. In connection, it would also be intriguing to study the effect of hydrodynamics and explicit solvent on this apparently universal dynamics of active polymer \cite{majumder2019,bera2022}.

\acknowledgments
This work was funded by the Deutsche Forschungsgemeinschaft (DFG, German Research Foundation) under Grant No.\ 189\,853\,844 -- SFB/TRR 102 (Project B04) and further supported by the Deutsch-Franz\"osische Hochschule (DFH-UFA) through the Doctoral College ``$\mathbb{L}^4$'' under Grant No.\ CDFA-02-07, and the Leipzig Graduate School of Natural Sciences ``BuildMoNa''. S.M. thanks the Science and Engineering Research Board (SERB), Govt. of India for a Ramanujan Fellowship (file no.\ RJF/2021/000044). S.P. acknowledges ICTS-TIFR, DAE, Govt. of India for a research fellowship. 

%
\end{document}


\title{Supplemental Material: Activity Induced Enhanced Diffusion of a Polymer in Poor Solvent}
\author{Suman Majumder}\email[]{ suman.jdv@gmail.com}
\affiliation{Amity Institute of Applied Sciences, Amity University Uttar Pradesh, Noida 201313,
India
}
\author{Subhajit Paul}\email[]{ subhajit.paul@icts.res.in}
\affiliation{International Center for Theoretical Sciences, Tata Institute of Fundamental Research, Bangalore-560089, India }
\author{Wolfhard Janke}\email[]{ wolfhard.janke@itp.uni-leipzig.de}
\affiliation{Institut f\"{u}r Theoretische Physik, Universit\"{a}t Leipzig, IPF 231101, 04081 Leipzig, Germany 
}
\date{\today}



\maketitle

\section{Model and Simulation Method}\label{model}
We consider a bead-spring model of a flexible polymer chain in which the monomers are connected in a linear way. The bonded interaction between successive monomers is modeled with the standard finitely extensible non-linear elastic (FENE) potential defined as
\begin{equation}\label{fene}
V_{\rm{FENE}}(r) = - \frac{K}{2}R^2 {\rm{ln}} \bigg[ 1- \bigg(\frac{r-r_0}{R}\bigg)^2\bigg]\,,
\end{equation}
where $K=40$ is the spring constant, $r_0=0.7$ the equilibrium bond length, and  $R=0.3$ is the maximum allowed extension of the bond. 
\par
The non-bonded interaction among different monomers with separation $r$ is modeled via the standard Lennard-Jones potential 
\begin{equation}
    V_{\rm{LJ}}(r) = 4\epsilon \bigg[\bigg(\frac{\sigma}{r}\bigg)^{12}- \bigg(\frac{\sigma}{r}\bigg)^6\bigg]\,,
\end{equation}
where $\epsilon=1$ is the interaction strength. The bead diameter $\sigma$ is related to $r_0$ as $\sigma= r_0/2^{1/6}$. This potential has a minimum at $2^{1/6}\sigma \equiv r_0$.
\par 
For computational benefit during simulations the LJ potential $V_{\rm{LJ}}$ is truncated and shifted at $r_c=2.5\sigma$ such that the non-bonded interaction has the form
\begin{equation}\label{nb_poten}
  V_{\rm{NB}}(r)=
\begin{cases}
  V_{\rm{LJ}}(r)-V_{\rm{LJ}}(r_c) -(r-r_c)\frac{dV_{\rm{LJ}}}{dr}\Big|_{r=r_c}  r<r_c \,,\\
0 ~~~~~~~~~ \text{otherwise}\,,
   \end{cases}
\end{equation}
which has the same qualitative behavior as $V_{\rm{LJ}}$. 
\par
Each bead is considered as an active Brownian particle. The activity for each bead works along its intrinsic propulsion direction which changes stochastically with time. Thus  the overdamped dynamics for each bead is modeled via the equations  in an implicit solvent
\begin{equation}\label{trans}
\partial_t{\vec{r}}_i = \frac{D_{\rm{tr}}}{k_BT} [f_p \hat{n}_i-\vec{\nabla} U_i]+\sqrt{2D_{\rm{tr}}}\,\vec{\Lambda}_i^{\rm{tr}},\\
\end{equation}
and
\begin{equation}\label{rot}
\partial_t{\hat{n}}_i = \sqrt{2D_{\rm{rot}}}(\hat{n}_i\times \vec{\Lambda}_i^{\rm{rot}}),
\end{equation}
where $\Vec{r}_i$ and $\hat{n}_i$ represents the position and orientational direction of the $i$-th bead, respectively, $U_i$ is the passive interaction consisting of both $V_{\rm{FENE}}$ and $V_{\rm{NB}}$, and $f_p$ denotes the strength of the self-propulsion force acting along $\hat{n}_i$. $\vec{\Lambda}_i^{\rm{tr}}$ and $\vec{\Lambda}_i^{\rm{rot}}$ are the random noises on the $i$-th bead with zero-mean and unit-variance and are Delta-correlated over different particles and time given by
\begin{equation}\label{noise_correl}
\langle\Lambda_{i}(t)\Lambda_{j}(t')\rangle = \delta_{ij} \delta(t,t')\,.
\end{equation}
In Eqs. \eqref{trans} and \eqref{rot} $D_{\rm{tr}}$ and $D_{\rm{rot}}$ are the translational and rotational diffusion constants which are related via the parameter $\Delta$ as
    \begin{equation}\label{delta}
	\Delta = \frac{D_{\rm{tr}}}{D_{\rm{rot}}\sigma^2}\,,
\end{equation}
where we have considered $\Delta=1/3$ in our simulations. Also $D_{\rm{tr}}=k_BT/ \gamma$, where  the drag coefficient $\gamma=1$. We have set the integration time step for MD simulations to $ 10^{-5}$ in units of the time scale  $\tau_0=\sigma^2\gamma/\epsilon$ ($\propto 1/D_{\rm{rot}}=\Delta \sigma^2 \gamma/k_BT$ at fixed $k_BT/\epsilon$).  
\par
The activity is measured in terms of a dimensionless quantity, the P\'eclet number $Pe$, defined as the ratio between the active force $f_p$ and the thermal force $k_BT/\sigma$ as 
\begin{equation}\label{pe_def}
 {Pe}=\frac{f_p\sigma}{k_BT}.
\end{equation}
In our simulations we choose the temperature $T=0.1\epsilon/k_B$, small enough to keep the thermal noise much lower compared to the active force. Such a choice of the temperature is well below the $\Theta$-transition temperature of the passive polymer which is the case with $Pe=0$. All the results in this Letter are presented in terms of $Pe$.

\par
The initial configurations are prepared at a high temperature where the polymer conformation is an extended coil. All our results for the mean square displacement (MSD), whether for the center of mass or a tagged monomer, are presented after the polymer reaches at its corresponding steady state.  We considered chains with $N$ varying between $32 \le N \le 380$ and self-propulsion value $Pe$ between $0 \le Pe \le 62.5$. 

\section{Trajectories for different $Pe$}
In Fig.\ \ref{trajectory} we show the typical trajectories for the (a) center of mass (cm), (b)  central bead (cb), and (c) end bead (eb) of a polymer of length $N=128$ for different values of $Pe$ at $T=0.1\epsilon/ k_B$. The trajectories presented are for a time period after the polymer reaches the steady state for the corresponding value of $Pe$. This was made certain by running the simulations for long enough time. As seen from Fig.\ 1 in the main text of the paper, the steady-state conformation of the polymer changes from globule to a coil with increasing activity. The corresponding ${\rm MSD}$s are calculated from these trajectories in the steady state.

\begin{figure*}[t!]\label{trajectory}
	\centering
	\includegraphics*[width=0.98\textwidth]{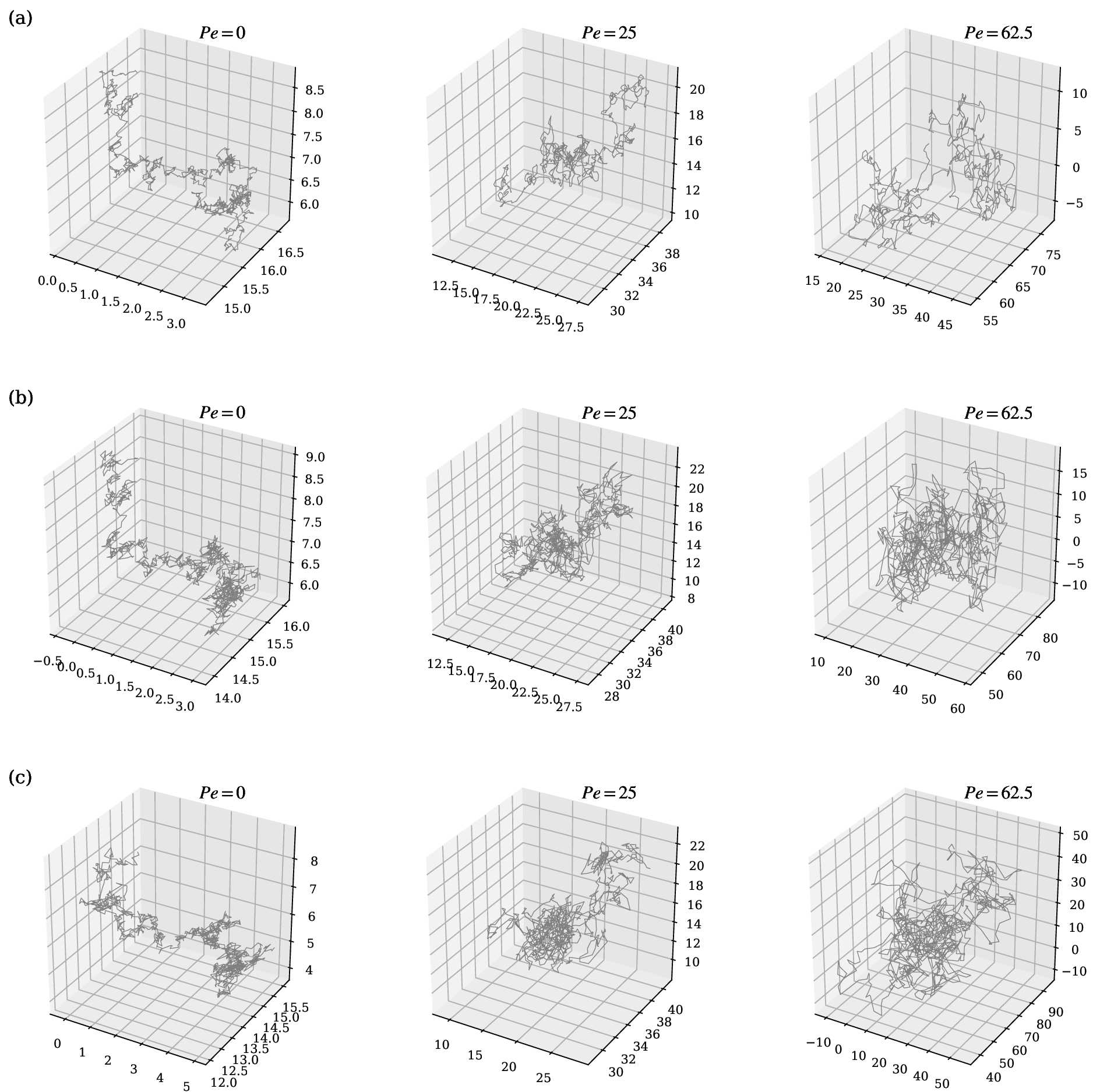}\\
	\caption{\label{trajectory} Typical trajectories of the (a) center of mass, (b) central monomer, and (c) end monomer of a polymer of length $N=128$, for different strengths of activity $Pe$ at temperature $T=0.1\epsilon/k_B$. Note the increasing spatial scales with increasing $Pe$.}
\end{figure*}

\section{Rouse model of active polymer}

To provide a theoretical understanding for the scaling behavior of our simulation data for the effective diffusion constant $D_{\rm eff}$ of  cm of the active polymer in poor solvent, we consider an analog of the Rouse model \cite{rouse1953}. For a passive polymer, the Rouse model correctly predicts the scaling of the diffusion constant in absence of hydrodynamic effects in good solvent. It is a simple bead-spring model where only a harmonic potential is considered to mimic the bonded interaction between successive monomers.  In the analog Rouse model, in addition to the above feature, we consider that the monomers are active Brownian particles, and there exists a non-bonded interaction between the monomers in order to take care of the poor solvent condition.  For a polymer chain of length $N$, the equation of motion for a bead which is not at the end can be written as:
\begin{eqnarray}\label{eom_rouse}
    \dot{\vec{r}}_i=-k(2\vec{r}_i-\vec{r}_{i-1}-\vec{r}_{i+1}) + \Vec{F}_i\,~~~~i\in[2,N-1],  
\end{eqnarray}
where $k$ is the spring constant of the harmonic bonds and $\Vec{F}_i$ is the net force acting on the bead due to a combination of the thermal noise, self propulsion or active force, and the non-bonded interaction in poor solvent condition. Since the end monomers experience only one bonded interaction, their equations of motion are given as
\begin{eqnarray}\label{eom1}
\dot{\vec{r}}_1=-k (\vec{r}_1-\vec{r}_2)  +\vec{F}_1, 
\end{eqnarray}
and 
\begin{eqnarray}\label{eomN}
\dot{\vec{r}}_N=-k (\vec{r}_N-\vec{r}_{N-1})  +\vec{F}_N.
\end{eqnarray}
Adding \eqref{eom_rouse}, \eqref{eom1}, and \eqref{eomN}, the equation of motion of the cm of the polymer is obtained as  
\begin{equation}\label{eq_com}
    \dot{\vec{r}}_{\rm cm} =\frac{1}{N}\sum_{i=1}^N \Vec{F}_i.
\end{equation}
Integrating Eq.\ \eqref{eq_com} we get
\begin{equation}
    \Vec{r}_{\text{cm}}(t) - \Vec{r}_{\text{cm}}(0)= \frac{1}{N}\int_0^t dt' \left[\sum_{i=1}^N \Vec{F}_i(t')\right].
\end{equation}
This leads to the expression for MSD of the cm of the polymer as
\begin{widetext}
\begin{eqnarray}\label{MSD_cm1}
\begin{split}
    \text{MSD}_{\text{cm}}(t) = \langle[\Vec{r}_{\text{cm}}(t) - \Vec{r}_{\text{cm}}(0)]^2\rangle = \frac{1}{N^2}\left \langle \int_0^t \int_0^t  dt' dt'' \left[\sum_{i=1}^N\Vec{F}_i(t') \right]\cdot \left[\sum_{j=1}^N\Vec{F}_j(t'')\right] \right\rangle.  
\end{split}
\end{eqnarray}
\end{widetext}
Now we assume that the net force $\vec{F}_i$ acting on an individual bead is random and delta-correlated over space and time, i.e., 
\begin{equation}\label{F_delta}
    \langle \Vec{F}_i(t') \Vec{F}_j(t'') \rangle = 6 \frac{D_{\rm a}}{D_m}\delta_{ij}\delta(t'-t'')\,,
\end{equation}
where $D_{\rm a}$ is the diffusion constant of an active particle and $D_m$ is the factor by which $D_{\rm a}$ gets modified in a poor solvent condition so that the effective diffusion constant of the bead becomes $D_{\rm a}/D_m$. Using Eq.\ \eqref{F_delta} in Eq.\ \eqref{MSD_cm1} we get 
\begin{equation}
 \text{MSD}_{\text{cm}}(t) = 6D_{\rm eff}t
\end{equation}
where $D_{\rm eff}$ is the effective diffusion constant of the cm of the polymer and is given as 
\begin{equation}\label{D_eff1}
 D_{\rm eff}=\frac{(D_{\rm a}/D_m)}{N}.
\end{equation}
 In Eq.\ \eqref{D_eff1}, the expression for $D_{\rm a}$ can be obtained from the MSD of an active Brownian particle as \cite{howse2007}
 \begin{widetext}
\begin{equation}\label{MSD_eq1}
\text{MSD}_{\rm a}(t)=	\langle \left[\vec{r}_i(t)-\vec{r}_i(0) \right]^2\rangle= 6D_{\rm tr}t+\frac{v_0^2}{D_{\rm rot}^2}\Big[ D_{\rm rot}t+\exp(-D_{\rm rot}t)-1\Big],
\end{equation}
\end{widetext}
where $D_{\text{tr}}$ and $D_{\text{rot}}$ are the translational and rotational diffusion constants, respectively, as defined previously in Sec.\ \ref{model}. At large $t~(\gg \tau_0)$, the rotational diffusion constant $D_{\rm rot} t \gg 1$ and Eq.\ \eqref{MSD_eq1} reduces to
\begin{equation}\label{MSD_eq2}
\text{MSD}_{\rm a}(t)= \left(6D_{\text{tr}}+\frac{v_0^2}{D_{\text{rot}}}\right)t\,.
\end{equation}
Inserting $D_{\rm tr}=k_BT/\gamma$, $D_{\rm rot}=3D_{\rm tr}/\sigma^2$, and the ballistic velocity for each bead $v_0 = f_p/\gamma=Pek_BT/\sigma\gamma$,  Eq.\ \eqref{MSD_eq2} transforms to
\begin{equation}\label{MSD_eq3}
 {\rm MSD}_{\rm a}(t) = \left(1+\frac{Pe^2}{18}\right)\frac{6k_BTt}{\gamma}.
\end{equation}
This provides the diffusion constant of an active particle as 
\begin{equation}
    D_{\rm a}=\left(1+\frac{Pe^2}{18}\right)\frac{k_BT}{\gamma}.
\end{equation}
Finally, inserting the above expression of $D_{\rm a}$ in Eq.\ \eqref{D_eff1} we get
\begin{equation}\label{D_eff2}
 D_{\rm eff}=\left(1+\frac{Pe^2}{18} \right)\frac{k_BT}{\gamma D_m N}.
\end{equation}

\section{Fitting results}
Equation\ \eqref{D_eff2} predicts the scaling laws $D_{\rm eff} \sim N^{-1}$ and $D_{\rm eff} \sim Pe^{2}$  for fixed $Pe$ and $N$, respectively. In order to verify the simulation data, however, one can assume that $D_{\rm eff} \sim N^{-x}$ and $D_{\rm eff} \sim Pe^{y}$ for fixed $Pe$ and $N$, respectively. Keeping Eq.\ \eqref{D_eff2} in mind, thus the ansatz for fitting can be written as
\begin{equation}\label{ansatz}
 D_{\rm eff}=\left(1+\frac{Pe^y}{18}\right)\frac{k_BT}{\gamma D_m N^x}.
\end{equation}
In Tables\ \ref{tab1}-\ref{tab4} we tabulate the results from our fitting exercise using the above ansatz for our simulation data of $D_{\rm eff}$, both as a function of $N$ and $Pe$.  
\begin{table}[ht!]
 \caption{Fitting results for different fixed $Pe$ using $y=2$ in the ansatz \eqref{ansatz} with both $D_m$ and $x$ as fit parameters. The quality of the fitting can be judged from the reduced chi-squared $\chi_r^2=\chi^2/{\rm d.o.f}$; where ${\rm d.o.f}$ is the number of degrees of freedom. From the results we conclude that the mean values of $D_m$ and $x$ are $0.47$ and $0.98$, respectively. The corresponding standard deviations are  $0.07$ and $0.03$, respectively.}
\begin{tabular}{m{1.0cm} m{1.25cm} m{1.25cm} m{0.7cm}} 
 \hline
 \hline
 $Pe$ & ~~$D_m$ & ~~~$x$ & $~\chi_r^2$ \\
 \hline
 \hline
  ~0.0& 0.46(4) & 0.96(2) &1.45\\  
  12.5& 0.41(8) & 1.00(4) &0.40\\ 
  25.0& 0.55(8) & 0.96(3) &0.52\\ 
  37.5& 0.37(7) & 1.02(4) &1.58\\ 
  50.0& 0.54(8) & 0.96(3) &2.17\\ 
  62.5& 0.49(8) & 0.96(5) &0.69\\ 
 \hline
\end{tabular}
\label{tab1}
\end{table}
\begin{table}[h!]
 \caption{Results from fitting for different fixed $Pe$ using $y=2$ in the ansatz \eqref{ansatz} with $D_m$ as the only fit parameter and fixing the exponent $x=1$. The obtained mean value of $D_m$ is $0.42$ with a standard deviation of $0.03$.}
\begin{tabular}{m{1.0cm} m{1.25cm} m{0.7cm}} 
 \hline
 \hline
 $Pe$ & ~~$D_m$ & $~\chi_r^2$ \\
 \hline
 \hline
  ~0.0& 0.39(4) &2.17\\  
  12.5& 0.41(1) &0.33\\ 
  25.0& 0.45(8) &0.64\\ 
  37.5& 0.42(7) &1.37\\ 
  50.0& 0.46(8) &2.13\\ 
  62.5& 0.40(8) &0.74\\ 
 \hline
\end{tabular}
\label{tab2}
\end{table}
\begin{table}[h!]
 \caption{Fitting results for different fixed $N$ using $x=1$ in the ansatz \eqref{ansatz} with both $D_m$ and $y$ as fit parameters. The obtained mean values of $D_m$ and $y$ are $0.39$ and $1.97$, respectively. The corresponding standard deviations are  $0.04$ and $0.03$, respectively.}
\begin{tabular}{m{1.0cm} m{1.25cm} m{1.25cm} m{0.7cm}} 
 \hline
 \hline
 ~$N$ & ~~$D_m$ & ~~~$y$ & $~\chi_r^2$ \\
 \hline
 \hline
  ~32& 0.47(4) & 2.00(2) &1.19\\  
  ~64& 0.40(2) & 1.98(2) &0.84\\ 
  128& 0.35(3) & 1.94(3) &0.98\\ 
  192& 0.37(3) & 1.94(2) &2.08\\ 
  256& 0.35(2) & 1.95(2) &2.94\\ 
  310& 0.40(3) & 2.00(2) &0.51\\ 
  380& 0.37(3) & 1.96(2) &0.45\\ 
 \hline
\end{tabular}
\label{tab3}
\end{table}
\begin{table}[h!]
 \caption{Results from fitting for different fixed $N$ using $x=1$ in the ansatz \eqref{ansatz} with $D_m$ as the only fit parameter and fixing the exponent $y=2$. The obtained mean value of $D_m$ is $0.42$ with a standard deviation of $0.03$.}
\begin{tabular}{m{1.0cm} m{1.25cm} m{0.7cm}} 
 \hline
 \hline
 ~$N$ & ~~$D_m$ & $~\chi_r^2$ \\
 \hline
 \hline
  ~32& 0.47(2) &0.96\\  
  ~64& 0.41(1) &0.77\\ 
  128& 0.42(2) &1.50\\ 
  192& 0.44(2) &2.80\\ 
  256& 0.40(3) &3.11\\ 
  310& 0.40(2) &0.41\\ 
  380& 0.42(1) &0.97\\
 \hline
\end{tabular}
\label{tab4}
\end{table}
 
From the presented fitting exercise we conclude that our simulation data is consistent with the prediction \eqref{D_eff2} using a Rouse model of active polymer. It also indicates that the only free parameter in the model $D_m$ appears to be robust, independent of the chain length $N$ and strength of activity $Pe$.
%
